\documentclass[12pt,a4paper]{article}

\usepackage{graphicx}
\usepackage{a4wide,cite}

\newcommand{\0}[1]{\mathbf{0}^{#1}}
\newcommand{\1}[1]{\mathbf{1}^{#1}}
\newcommand{\2}[1]{\mathbf{2}^{#1}}
\newcommand{\Li}[1]{\mathop{\mathrm{Li}}\nolimits_{#1}}

\begin{document}

 \thispagestyle{empty}
 \begin{flushright}
 {Budker INP 2001-10} \\[3mm]
 {MZ-TH/01--01} \\[3mm]
 {hep-ph/0103078} \\[3mm]
 {March 2001}
 \end{flushright}
 \vspace*{2.0cm}
 \begin{center}
 {\Large \bf
HQET quark-gluon vertex at one loop}
 \end{center}
 \vspace{1cm}  
 \begin{center}
 A.I.~Davydychev$^{a,}$\footnote{On leave from
 Institute for Nuclear Physics, Moscow State University,
 119899, Moscow, Russia. Email:
 davyd@thep.physik.uni-mainz.de} \ \ and \ \
 A.~G.~Grozin$^{b,}$\footnote{Email: A.G.Grozin@inp.nsk.su}
\\
 \vspace{1cm}
$^{a}${\em
 Department of Physics,
 University of Mainz, \\
 Staudinger Weg 7,
 D-55099 Mainz, Germany}
\\
\vspace{.3cm}  
$^{b}${\em 
Budker Institute of Nuclear Physics, \\
Novosibirsk 630090, Russia}
\\
\end{center}
 \hspace{3in}
 \begin{abstract}
We calculate the HQET quark-gluon vertex at one loop,
for arbitrary external momenta,
in an arbitrary covariant gauge and space-time dimension.
Relevant results and algorithms for the three-point HQET integrals 
are presented.
We also show how one can obtain the HQET limit directly from
QCD results for the quark-gluon vertex.
 \end{abstract}

%%%%%%%%%%%%%%%%%%%%%%%%%%%%%%%%%%%%%%%%%%%%%%%%%%%%%%%%%%%%%%%%%%%%%%%
\newpage

\section{Introduction}
\label{SecWard}

Heavy Quark Effective Theory (HQET) is an effective field theory
approximating QCD for problems with a single heavy quark having mass $m$,
when characteristic momenta of light fields are much lower than $m$,
and there exists a 4-velocity $v$ such that characteristic residual momenta
$k=p-mv$ of the heavy quark are also small.
It has substantially improved our understanding of heavy quark physics
during the last decade~\cite{N,MW}.
Methods of perturbative calculations in HQET are reviewed in~\cite{Lect}.

In this paper, we calculate the quark-gluon vertex
in the leading-order HQET ($1/m^0$) at one loop,
for arbitrary external momenta, in an arbitrary covariant gauge,
in space-time dimension $d=4-2\varepsilon$.
This allows us to take all on-shell limits
(introducing additional $1/\varepsilon$ divergences) directly.
The general $d$-dimensional results can also be used
for expansion around a dimension other than 4;
for example, 2-dimensional HQET was considered in the literature
in some detail.

A one-loop calculation of the QCD quark-gluon vertex
with a finite quark mass $m$ has been recently completed~\cite{DOS}
(where references to earlier partial results can be found).
We check how the HQET result can be obtained by taking the limit
$m\to\infty$ in the QCD result.

\begin{figure}[b]
\begin{center}
\begin{picture}(29,19)
\put(14.5,9.5){\makebox(0,0){\includegraphics{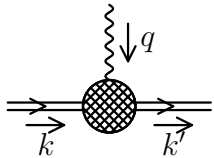}}}
\put(6,-2){\makebox(0,0)[b]{$k$}}
\put(23,-2){\makebox(0,0)[b]{$k'$}}
\put(18.5,13){\makebox(0,0)[l]{$q$}}
\end{picture}
\end{center}
\caption{HQET quark-gluon vertex}
\label{FigVert}
\end{figure}

Let the sum of bare one-particle-irreducible vertex diagrams in HQET
(Fig.~\ref{FigVert}) be \ $\mathrm{i} g_0 t^a \Gamma^\mu(k,q)$.
The ``full'' momenta of the incoming quark and the outgoing one are
\begin{equation}
p = mv + k,\quad
p' = mv + k',
\label{ResMom}
\end{equation}
where $v$ is the heavy-quark 4-velocity ($v^2=1$),
and $k$, $k'$ are the residual momenta.
The momentum transfer is $q=p'-p=k'-k$.
In the HQET limit, $m\to\infty$,
$k\sim k'\sim q\sim\mathcal{O}(1)$.
The heavy quark propagator in HQET is
\begin{equation}
S(k) = \frac{\rlap/v+1}{2} \frac{1}{k\cdot v+\mathrm{i}0},
\end{equation}
and the elementary quark-gluon vertex is $\mathrm{i} g_0 t^a v^\mu$.

In the leading-order HQET, heavy-quark propagators and vertices
do not depend on the component of the heavy-quark momenta
orthogonal to $v$.
Therefore, $\Gamma^\mu(k,q)$ does not depend on $k_\bot=k-(k\cdot v)v$.
The only vectors in the problem are $v$ and $q$,
and (see~\cite{Lect})
\begin{equation}
\Gamma^\mu(k,q) = \Gamma_v(\omega,\omega',q^2) v^\mu
+ \Gamma_q(\omega,\omega',q^2) q^\mu,
\label{VertStruct}
\end{equation}
where
\begin{equation}
\omega \equiv k\cdot v,\quad
\omega' \equiv k'\cdot v
\label{ResEn}
\end{equation}
are the residual energies, and $q\cdot v=\omega'-\omega$.
The functions $\Gamma_v$ and $\Gamma_q$ can be reconstructed
from the contractions
\begin{equation}
\Gamma_v = \frac{(\omega'-\omega)\Gamma^\mu q_\mu-q^2\Gamma^\mu v_\mu}{Q^2},
\quad \quad
\Gamma_q = \frac{(\omega'-\omega)\Gamma^\mu v_\mu-\Gamma^\mu q_\mu}{Q^2},
\label{VertCoeff}
\end{equation}
where
\begin{equation}
Q^2 \equiv (\omega'-\omega)^2 - q^2
\label{Q2}
\end{equation}
is the 3-momentum transfer squared in the $v$ rest frame.

\begin{figure}
\begin{center}
\begin{picture}(60,22)
\put(30,11){\makebox(0,0){\includegraphics{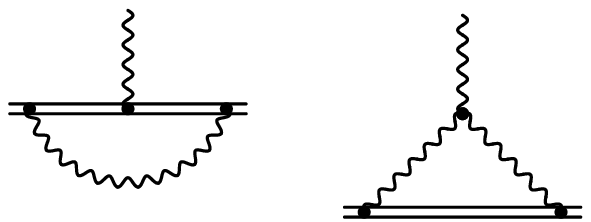}}}
\put(13,-2.5){\makebox(0,0)[b]{a}}
\put(47,-2.5){\makebox(0,0)[b]{b}}
\end{picture}
\end{center}
\caption{One-loop vertex diagrams}
\label{FigVert1}
\end{figure}

At the tree level, $\Gamma^\mu=v^\mu$.
One-loop corrections are shown in Fig.~\ref{FigVert1}.
The contribution $\Gamma_a^\mu$ of the diagram Fig.~\ref{FigVert1}a
is proportional to $v^\mu$;
that of Fig.~\ref{FigVert1}b has both structures.

\begin{figure*}[b]
\begin{center}
\begin{picture}(167,34)
\put(83.5,18){\makebox(0,0){\includegraphics{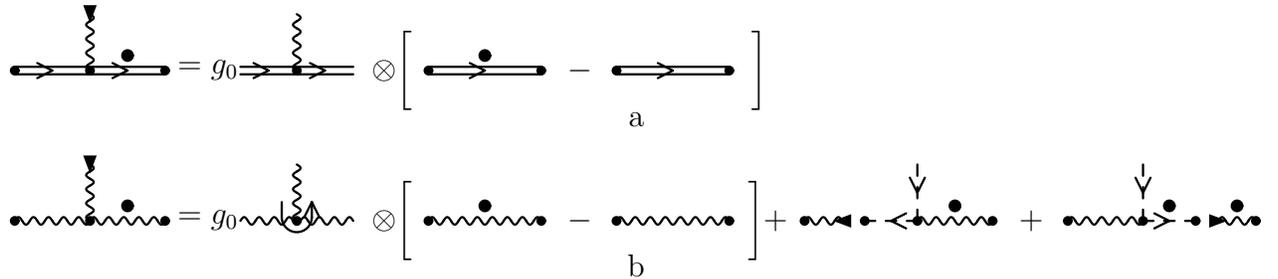}}}
\put(26,25){\makebox(0,0){${}=g_0$}}
\put(51,25){\makebox(0,0){$\otimes\Biggl[\Biggr.$}}
\put(76,25){\makebox(0,0){$-$}}
\put(101.5,25){\makebox(0,0){$\Biggl.\Biggr]\phantom{+}$}}
\put(26,5){\makebox(0,0){${}=g_0$}}
\put(51,5){\makebox(0,0){$\otimes\Biggl[\Biggr.$}}
\put(76,5){\makebox(0,0){$-$}}
\put(101,5){\makebox(0,0){$\Biggl.\Biggr]+$}}
\put(136,5){\makebox(0,0){$+$}}
\put(83.5,17.5){\makebox(0,0)[b]{a}}
\put(83.5,-2.5){\makebox(0,0)[b]{b}}
\end{picture}
\end{center}
\caption{Longitudinal gluon insertions into quark and gluon propagators}
\label{FigWardQG}
\end{figure*}

The contraction $\Gamma^\mu(k,q)q_\mu$ can be simplified
using the identities shown in Fig.~\ref{FigWardQG}.
Here a gluon line with a black triangle at the end denotes
a ``longitudinal gluon insertion'';
when attached to a vertex,
it means just the contraction
with the incoming gluon momentum 
(note that it contains no gluon propagator!).
A dot near a propagator means that its momentum is shifted by $q$.
The colour structures are singled out as prefactors
in front of the propagator differences.
The circular arrow in Fig.~\ref{FigWardQG}b shows the order of indices
in the colour structure of the three-gluon vertex ${\rm i}f^{abc}$.
Two last terms in Fig.~\ref{FigWardQG}b contain
longitudinal gluon insertions again;
for them, the identities of Fig.~\ref{FigWardQG} can be recursively used.

\begin{figure*}
\begin{center}
\begin{picture}(146,77)
\put(73,38.5){\makebox(0,0){\includegraphics{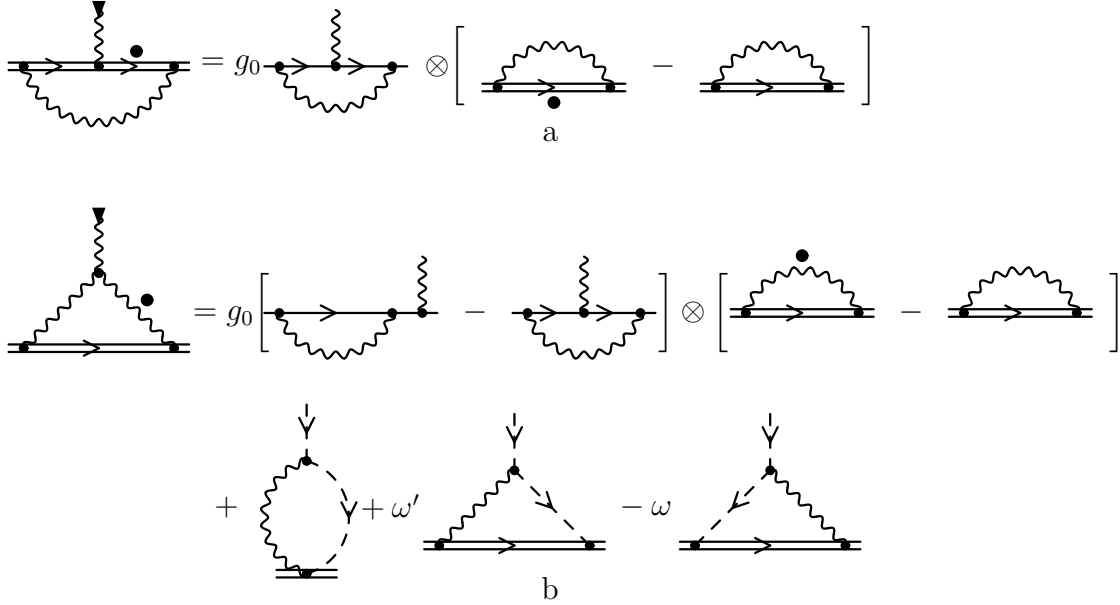}}}
\put(30,68.5){\makebox(0,0){${}=g_0$}}
\put(59,68.5){\makebox(0,0){$\otimes\Biggl[\Biggr.$}}
\put(88,68.5){\makebox(0,0){$-$}}
\put(117,68.5){\makebox(0,0){$\Biggl.\Biggr]\phantom{\otimes}$}}
\put(30,35.6875){\makebox(0,0){${}=g_0\Biggl[\Biggr.$}}
\put(63,35.6875){\makebox(0,0){$-$}}
\put(92,35.6875){\makebox(0,0){$\Biggl.\Biggr]\otimes\Biggl[\Biggr.$}}
\put(121,35.6875){\makebox(0,0){$-$}}
\put(150,35.6875){\makebox(0,0){$\Biggl.\Biggr]\phantom{\otimes}$}}
\put(30,9.75){\makebox(0,0){$+$}}
\put(51.25,9.75){\makebox(0,0){${}+\omega'$}}
\put(85.25,9.75){\makebox(0,0){${}-\omega$}}
\put(73,58.5){\makebox(0,0)[b]{a}}
\put(73,-2.5){\makebox(0,0)[b]{b}}
\end{picture}
\end{center}
\caption{Ward--Slavnov--Taylor identities for the one-loop vertex}
\label{FigWard}
\end{figure*}

\begin{figure}
\begin{center}
\begin{picture}(47,12)
\put(23.5,5){\makebox(0,0){\includegraphics{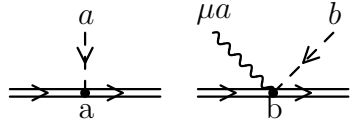}}}
\put(11,10){\makebox(0,0)[b]{$a$}}
\put(28,10){\makebox(0,0)[b]{$\mu a$}}
\put(44,10){\makebox(0,0)[b]{$b$}}
\put(11,-2.5){\makebox(0,0)[b]{a}}
\put(36,-2.5){\makebox(0,0)[b]{b}}
\end{picture}
\end{center}
\caption{Non-standard vertices}
\label{FigNS}
\end{figure}

\begin{figure}
\begin{center}
\begin{picture}(52,32)
\put(26,16){\makebox(0,0){\includegraphics{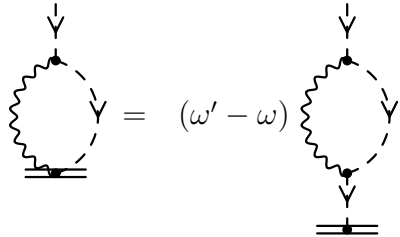}}}
\put(25,16){\makebox(0,0){$\hspace{3mm}=\hspace{3mm}(\omega'-\omega)$}}
\end{picture}
\end{center}
\caption{Relation between ghost self-energy contributions}
\label{FigRel}
\end{figure}

Application of these identities to the diagrams of Fig.~\ref{FigVert1}
is shown in Fig.~\ref{FigWard}.
Here the non-standard vertices of Fig.~\ref{FigNS} are
$\mathrm{i} g_0 t^a$ and $g_0^2 f^{abc} t^c v^\mu$.
This is, of course, just the one-loop case of
the general Ward--Slavnov--Taylor identity for the quark-gluon vertex,
which is discussed in~\cite{PT-QCD,DOS} in detail.
The one-loop contributions to this identity are collected
in Eqs.~(2.28) and (2.29) of~\cite{DOS}.
They can be easily associated
with the diagrams shown in Fig.~\ref{FigWard}. The only term
which requires some explanation is the diagram involving
one-loop ghost self-energy contribution (the first diagram 
in the second line of Fig.~\ref{FigWard}b). 
Its connection with the last contribution on the r.h.s.\ of~Eq.~(2.29)
of~\cite{DOS} is shown in Fig.~\ref{FigRel}. This relation can
be easily verified by direct calculation.
Using Fig.~\ref{FigRel}, we can avoid introducing the 
non-standard vertex shown in~Fig.~\ref{FigNS}b.

\begin{figure}
\begin{center}
\begin{picture}(32,10)
\put(16,5){\makebox(0,0){\includegraphics{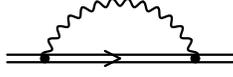}}}
\end{picture}
\end{center}
\caption{Heavy-quark self energy}
\label{FigSigma}
\end{figure}

For the contributions of the diagrams of Fig.~\ref{FigVert1}a,b,
we have
\begin{eqnarray}
\Gamma_a^\mu(k,q) q_\mu &=& \left( 1 - \frac{C_A}{2 C_F} \right)
\left( \Sigma(\omega) - \Sigma(\omega') \right),
\nonumber\\
\Gamma_b^\mu(k,q) q_\mu &=& \frac{C_A}{2 C_F}
\left( \Sigma(\omega) - \Sigma(\omega') \right)
+ (\mbox{ghost terms}).\hspace{3mm}
\label{WST}
\end{eqnarray}
Here $-\mathrm{i}\Sigma(\omega)$ is given by the one-loop self-energy
diagram of Fig.~\ref{FigSigma}:
\begin{eqnarray}
\Sigma(\omega) &=& \frac{C_F}{2}\; \frac{g_0^2}{(4\pi)^{d/2}} 
\left(2+(d-3)\xi\right) \mathcal{I}(\omega),
\label{Sigma}\\
\mathcal{I}(\omega) &=& - \frac{\mathrm{i}}{\pi^{d/2}} \int
\frac{\mathrm{d}^d l}{(l\cdot v+\omega+\mathrm{i}0)(l^2+\mathrm{i}0)}
\nonumber\\
&=& 2 (-2\omega)^{d-3} \Gamma(3-d)\Gamma(d/2-1)
\label{Idef}
\end{eqnarray}
(see~\cite{BG}).
In what follows, we shall not explicitly write $+\mathrm{i}0$
in denominators.
Here $\xi=1-a_0$, $a_0$ is the bare gauge-fixing parameter.
We can see that the Yennie gauge~\cite{FY} (see also in
\cite{AbrSol}) is of special interest, since $\Sigma(\omega)$
is finite at $\xi=-2$. Moreover, if the generalization
of Yennie gauge to an arbitrary dimension is chosen as
$\xi=-2/(d-3)$~\cite{FYgen}, then 
in the Abelian case (and, in particular, at one loop),
the heavy-quark self energy vanishes~\cite{BG} 
(see~\cite{Lect} for a tutorial).
The two-loop HQET self energy was obtained in~\cite{BG};
the three-loop one can be calculated using the methods of~\cite{Gr}.
These calculations are based on integration by parts~\cite{CT}.

Using the identity~(\ref{WST}), we can obtain the result
for the diagram of Fig.~\ref{FigVert1}a without calculations:
\begin{equation}
\Gamma_a^\mu(k,q) = - \left(1 - \frac{C_A}{2 C_F}\right)
\frac{\Sigma(\omega')-\Sigma(\omega)}{\omega'-\omega}
v^\mu.
\label{Vert1a}
\end{equation}
This result is also confirmed by direct calculation.
Feynman integrals of the type of Fig.~\ref{FigVert1}a,
\begin{equation}
\label{Int_a}
\int \frac{\mathrm{d}^d l}
{(l\cdot v+\omega)^{\nu_1}(l\cdot v+\omega')^{\nu_2}(l^2)^{\nu_3}},
\end{equation}
can always be calculated, for integer $\nu_1$, $\nu_2$,
by applying
\[
\frac{1}{(l\cdot v+\omega)(l\cdot v+\omega')} 
= \frac{1}{\omega'-\omega} \left[
\frac{1}{l\cdot v+\omega}-\frac{1}{l\cdot v+\omega'}
\right]
\]
required number of times. We note that in~\cite{Bouzas} integrals
of the type (\ref{Int_a}) with different velocities $v$ have
been examined.

Calculation of Fig.~\ref{FigVert1}b requires
more complicated Feynman integrals.
A method of their calculation is presented in Sect.~\ref{SecInt}.
Results for the HQET vertex, as well as various limiting cases,
are discussed in Sect.~\ref{SecVert}.
Sect.~\ref{Concl} contains a brief summary of the results obtained.
In Appendix~\ref{AppInt}, we present general results
for the Feynman integrals of the type of Fig.~\ref{FigVert1}b,
for arbitrary $d$ and powers of the denominators.
In Appendix~\ref{AppThreshold} some issues related to the
$m\to\infty$ limit of scalar integrals occurring in QCD are examined.
In Appendix~\ref{AppQCD}, we discuss the relation between
the QCD vertex at $k,q\ll m$ and the HQET vertex.
In Appendix~\ref{AppBG}, we present the HQET vertex for
the heavy-quark scattering in an external gluon field
in the background-field formalism.

\section{Triangle integrals}
\label{SecInt}

\subsection{Recurrence relations}
\label{SubRec}

\begin{figure}
\begin{center}
\begin{picture}(42,26)
\put(21,13){\makebox(0,0){\includegraphics{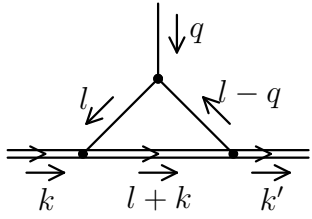}}}
\put(6,-2.5){\makebox(0,0)[b]{$k$}}
\put(21,-2.5){\makebox(0,0)[b]{$l+k$}}
\put(36,-2.5){\makebox(0,0)[b]{$k'$}}
\put(11,11){\makebox(0,0)[b]{$l$}}
\put(33,11){\makebox(0,0)[b]{$l-q$}}
\put(25,21){\makebox(0,0)[l]{$q$}}
\end{picture}
\end{center}
\caption{Feynman integral $V$}
\label{FigV}
\end{figure}

We consider the class of Feynman integrals (Fig.~\ref{FigV})
\begin{equation}
V(\nu_0,\nu_1,\nu_2) = - \frac{\mathrm{i}}{\pi^{d/2}} \int
\frac{\mathrm{d}^d l}
     {(l\cdot v+\omega)^{\nu_0}(l^2)^{\nu_1}((l-q)^2)^{\nu_2}}.
\label{Vdef}
\end{equation}
The cases $\nu_0=0$, $\nu_2=0$, $\nu_1=0$ are trivial,
the results are proportional to
\begin{eqnarray}
V(0,1,1) &=& \mathcal{G}(q^2)
= (-q^2)^{d/2-2} \frac{\Gamma(2-d/2)\Gamma^2(d/2-1)}{\Gamma(d-2)},
\nonumber\\
V(1,1,0) &=& \mathcal{I}(\omega),\quad
V(1,0,1) = \mathcal{I}(\omega').
\label{triv}
\end{eqnarray}

When all the indices are non-zero, we use integration by parts~\cite{CT},
similarly to~\cite{Da,BG}.
Applying the operators $(\partial/\partial l)\cdot v$,
$(\partial/\partial l)\cdot l$, and $(\partial/\partial l)\cdot(l-q)$
to the integrand of~(\ref{Vdef}), we obtain the recurrence relations
\begin{eqnarray}
\left[\nu_0\0+ + 2\nu_1\1+(\0--\omega) + 2\nu_2\2+(\0--\omega')\right] V &=& 0,
\hspace{6mm}
\label{r0}\\
\left[d-\nu_0-\nu_2-2\nu_1 + \omega\; \nu_0\0+ + \nu_2\2+(q^2-\1-)\right] V &=& 0,
\hspace{6mm}
\label{r1}\\
\left[d-\nu_0-\nu_1-2\nu_2 + \omega'\nu_0\0+ + \nu_1\1+(q^2-\2-)\right] V &=& 0,
\hspace{6mm}
\label{r2}
\end{eqnarray}
where $\0\pm V(\nu_0,\nu_1,\nu_2)=V(\nu_0\pm1,\nu_1,\nu_2)$, etc.
When constructing the recurrence procedure we assume that all indices
$\nu_i$ are integer.

If $\nu_1<0$ and $\nu_2\ne1$, we can raise $\nu_1$ by~(\ref{r1});
if $\nu_1<0$ and $\nu_2=1$, $\nu_0\ne1$, we can raise $\nu_1$ 
or $\nu_2$ by~(\ref{r0});
if $\nu_1<0$ and $\nu_2=\nu_0=1$, we can raise $\nu_1$ or $\nu_0$ by~(\ref{r2}).
The case $\nu_2<0$ is symmetric.
If $\nu_1>1$, we can lower it or $\nu_2$ by~(\ref{r2});
the case $\nu_2>1$ is symmetric.
We are left with $V(\nu_0,1,1)$.

Let us take $q^2/2$ times~(\ref{r0}),
add $\omega$ times~(\ref{r2}) and $\omega'$ times~(\ref{r1}),
and subtract the $\0-$ shifted sum of~(\ref{r1}) and (\ref{r2}).
We obtain at $\nu_1=\nu_2=1$
\begin{eqnarray}
&& \hspace*{-3mm}
\biggl[\frac{\Omega}{2}\nu_0\0+ \!+\! (d\!-\!2\nu_0\!-\!2)(\omega\!+\!\omega')
\!-\!2(d\!-\!\nu_0\!-\!2)\0- \biggr] V(\nu_0,1,1)
\nonumber\\
&&{} = \left[\1+\2-(\omega-\0-) + \2+\1-(\omega'-\0-)\right] V(\nu_0,1,1),
\hspace{5mm}
\label{r3}
\end{eqnarray}
where
\begin{equation}
\Omega \equiv q^2 + 4\omega\omega' = (\omega'+\omega)^2 - Q^2.
\label{Om}
\end{equation}
The integrals on the right-hand side of~(\ref{r3}) are trivial.
This relation allows us to raise or lower $\nu_0$.

Therefore, all integrals~(\ref{Vdef}) can be expressed, exactly at any $d$,
as linear combinations of three trivial integrals (\ref{triv})
and a non-trivial one, $V(1,1,1)\equiv\mathcal{V}(\omega,\omega',Q)$.
An implementation of this algorithm in {\sf REDUCE}
can be obtained at\\
http://wwwthep.physik.uni-mainz.de/Publications/progdata/mzth0101/ .

\subsection{Master integral}
\label{SubMaster}

The master integral
\begin{equation}
\mathcal{V}(\omega,\omega',Q) = - \frac{\mathrm{i}}{\pi^{d/2}} \int
\frac{\mathrm{d}^d l}{(l\cdot v+\omega)l^2(l-q)^2}
\label{V0def}
\end{equation}
is convergent at $d=4$, except for the case $q^2=0$
when it has a collinear divergence (infrared if $q=0$).
We shall consider the region $\omega<0$, $\omega'<0$, $q^2<0$,
where no real intermediate states exist.

Using the HQET version of Feynman parametrization (see, e.g., \cite{Lect})
\begin{equation}
\frac{1}{a^\alpha b^\beta} =
\frac{\Gamma(\alpha+\beta)}{\Gamma(\alpha)\Gamma(\beta)}
\int\limits_0^\infty \frac{y^{\beta-1}\mathrm{d}y}{(a+by)^{\alpha+\beta}}
\label{Feynman}
\end{equation}
twice, we have
\begin{eqnarray}
\mathcal{V} &=& \frac{4\mathrm{i}}{\pi^{d/2}} \int
\frac{\mathrm{d}^d l\,\mathrm{d}y\,\mathrm{d}y'}
{\left[-yl^2-y'(l-q)^2- 2 l\cdot v- 2 \omega\right]^3}
\nonumber\\
&=& -2 \Gamma(1+\varepsilon)
\int\limits_0^\infty \int\limits_0^\infty
\frac{\mathrm{d}y\,\mathrm{d}y'}{(y+y')^{1-2\varepsilon}
\left[1-2(\omega y+\omega'y')-q^2yy'\right]^{1+\varepsilon}},
\hspace*{6mm}
\label{V0a}
\end{eqnarray}
where $y$ and $y'$ have the dimensionality of inverse energy.

For $\varepsilon=0$ ($d=4$),
calculating the integral in $y'$ and substituting $y=1/z$, we obtain
\begin{equation}
\mathcal{V} = 2 \int\limits_0^\infty \frac{\log\frac{-q^2-2\omega'z}{z(z-2\omega)}\,
\mathrm{d}z}{(z+Q-\omega+\omega')(z-Q-\omega+\omega')},
\label{V0b}
\end{equation}
where $z$ has the dimensionality of energy.
Separating the logarithm as
\[
\log\frac{-q^2-2\omega'z}{z(z-2\omega)} =
- \log\frac{z-2\omega'}{Q-\omega-\omega'}
+ \log\frac{-q^2-2\omega'z}{z(Q-\omega-\omega')}
\]
and making the substitution $z=(-q^2)/z'$ in the second integral,
we obtain the representation
\begin{eqnarray}
\mathcal{V} &=&{} - 2 \int\limits_0^\infty
\frac{\log\frac{z-2\omega}{Q-\omega-\omega'}\,
\mathrm{d}z}{(z+Q-\omega+\omega')(z-Q-\omega+\omega')}
\nonumber\\
&&{} - 2 \int\limits_0^\infty
\frac{\log\frac{z-2\omega'}{Q-\omega-\omega'}\,
\mathrm{d}z}{(z+Q+\omega-\omega')(z-Q+\omega-\omega')},
\label{V0c}
\end{eqnarray}
which is explicitly symmetric under $\omega\leftrightarrow\omega'$.
The poles of the denominators at $z=Q\pm(\omega-\omega')$
are compensated by the corresponding zeros of the numerators.

Finally, we obtain
\begin{eqnarray}
Q \mathcal{V}(\omega,\omega',Q)
&=& \Li2\left(\frac{Q+\omega+\omega'}{2\omega}\right)
+ \Li2\left(\frac{Q+\omega+\omega'}{2\omega'}\right)
\nonumber\\
&& \hspace*{-4mm} 
+ \Li2\left(\frac{Q+\omega-\omega'}{2\omega}\right)
+ \Li2\left(\frac{Q-\omega+\omega'}{2\omega'}\right)
\nonumber\\
&& \hspace*{-4mm} 
+ \log\frac{Q-\omega-\omega'}{-2\omega}
\log\frac{Q-\omega+\omega'}{-2\omega}
\nonumber\\
&& \hspace*{-4mm}
+ \log\frac{Q-\omega-\omega'}{-2\omega'}
\log\frac{Q+\omega-\omega'}{-2\omega'}
- \frac{\pi^2}{3}.
\hspace*{6mm}
\label{V0res}
\end{eqnarray}
This expression has cuts at $\omega>0$, $\omega'>0$, and $Q<|\omega'-\omega|$,
where real intermediate states exist.

\section{HQET quark-gluon vertex}
\label{SecVert}

\subsection{General results}
\label{SubGen}

Here we present the one-loop HQET vertex, for arbitrary $d$ and $\xi$.
The contribution of the diagram of Fig.~\ref{FigVert1}a
was given in~(\ref{Vert1a}).
For Fig.~\ref{FigVert1}b, we obtain
\begin{eqnarray}
&& \hspace*{-10mm}
\Gamma_b^\mu q_\mu = \frac{C_A}{2 C_F}
\left(\Sigma(\omega)-\Sigma(\omega')\right)
- C_A \frac{g_0^2}{(4\pi)^{d/2}} \frac{\omega'-\omega}{4\Omega}
\nonumber\\
&&{}\times\biggl\{ 2 (\omega'+\omega)
\left[q^2+\omega\omega'(4+(d-4)\xi)\right] \mathcal{V}(\omega,\omega',Q)
\nonumber\\
&&\quad{} - \left[q^2(4-\xi)+4\omega\omega'(4+(d-4)\xi)\right] \mathcal{G}(q^2)
\nonumber\\
&&\quad{} + (d-3)\xi 
\left(\omega'\mathcal{I}(\omega)+\omega \mathcal{I}(\omega')\right)
\biggr\},
\label{Gq}\\
&& \hspace*{-10mm}
\Gamma_b^\mu v_\mu = C_A \frac{g_0^2}{(4\pi)^{d/2}}
\frac{\xi}{8 q^2 \Omega^2}
\biggl\{ - 2 (d-4) (\omega'+\omega)
\nonumber\\
&&\quad{}\times
\Bigl[ (d-6) \xi Q^2 q^2 \omega \omega'
+ \Omega Q^2 \bigl(2q^2 + (4+(d-4)\xi)\omega\omega'\bigr)
\nonumber\\
&&\qquad{} + (2+(d-3)\xi) \Omega q^2 \omega \omega' \Bigr]
\mathcal{V}(\omega,\omega',Q)
\nonumber\\
&&\quad{} + \Bigl[ 4 (d-3) (d-6) \xi q^2 \omega \omega' (\omega'+\omega)^2
\nonumber\\
&&\qquad{} - 2 \Omega q^2 \bigl(q^2 - 4\omega\omega'(d-4+(d-3)\xi)\bigr)
\nonumber\\
&&\qquad{} + \Omega Q^2 \bigl((d-4)(4+(d-4)\xi)\Omega
\nonumber\\
&&\qquad\quad{} + (d-3)(4-(d-4)\xi)q^2\bigr) \Bigr] \mathcal{G}(q^2)
\nonumber\\
&&\quad{} - (d-3) (d-6) \xi q^2 (\omega'+\omega)^2
\bigl(\omega \mathcal{I}(\omega') + \omega' \mathcal{I}(\omega)\bigr)
\nonumber\\
&&\quad{} - (d-3) \Omega \Bigl[
2 q^2 (\omega'-\omega) \bigl(\mathcal{I}(\omega')-\mathcal{I}(\omega)\bigr)
\nonumber\\
&&\qquad{} + (4+(d-4)\xi) (\omega^{\prime2}-\omega^2)
\bigl(\omega \mathcal{I}(\omega') - \omega' \mathcal{I}(\omega)\bigr)
\nonumber\\
&&\qquad{} + 2 (3+\xi) q^2 
\bigl(\omega \mathcal{I}(\omega') + \omega' \mathcal{I}(\omega)\bigr)
\Bigr]\biggr\},
\label{Gv}
\end{eqnarray}
where $\Omega$ is defined in~(\ref{Om}).
We have also derived the first contraction using~(\ref{WST})
(Fig.~\ref{FigWard}).
The second contraction vanishes in the Feynman gauge,
because the three-gluon vertex yields 0 when contracted with $v$ in all 3 indices.
We also note that there are some cancellations in the generalized
Yennie gauge, $\xi=-2/(d-3)$, as well as in the ``singular'' gauge $\xi=-4/(d-4)$
(which was discussed in~\cite{DOT1} in connection with the three-gluon vertex).

When the bare vertex $\Gamma^\mu$ is expressed via the renormalized quantities
\[
\frac{g_0^2}{(4\pi)^{d/2}} = \frac{\alpha_{\mathrm{s}}}{4\pi} \mu^{2\varepsilon}
%\Gamma(1-\varepsilon) 
e^{\gamma\varepsilon}
\left[1 + \mathcal{O}(\alpha_{\mathrm{s}})\right],\quad
a_0 = a \left[1 + \mathcal{O}(\alpha_{\mathrm{s}})\right],
\]
it should become $Z_\Gamma \Gamma_{\mathrm{r}}^\mu$,
where $Z_\Gamma=1+Z_1\alpha_{\mathrm{s}}/(4\pi\varepsilon)+\cdots$
is a minimal renormalization constant,
and the renormalized vertex $\Gamma_{\mathrm{r}}^\mu$ is finite
in the limit $\varepsilon\to0$.
Retaining only the pole parts $\mathcal{I}(\omega)\to2\omega/\varepsilon$,
$\mathcal{G}(q^2)\to1/\varepsilon$, $\mathcal{V}(\omega,\omega',Q)\to0$,
we obtain, either from~(\ref{Gq}) or from~(\ref{Gv}),
\begin{equation}
Z_\Gamma = 1 + \left[(a-3) C_F + \frac{a+3}{4} C_A\right]
\frac{\alpha_{\mathrm{s}}}{4\pi\varepsilon}
\label{ZG}
\end{equation}
When $g_0\Gamma^\mu=g\Gamma_{\mathrm{r}}^\mu Z_\alpha^{1/2}Z_\Gamma$
is multiplied by the external leg renormalization factors $Z_Q Z_A^{1/2}$,
it should give a finite matrix element.
Using
\begin{eqnarray*}
Z_A &=& 1 - \left[\frac{1}{2}\left(a-\frac{13}{3}\right)
+ \frac{4}{3} T_F n_l\right]
\frac{\alpha_{\mathrm{s}}}{4\pi\varepsilon},\\
Z_Q &=& 1 - (a-3) C_F \frac{\alpha_{\mathrm{s}}}{4\pi\varepsilon}
\end{eqnarray*}
($Z_Q$ follows from~(\ref{Sigma}), $n_l$ is the number of light flavours),
we arrive at
\begin{equation}
Z_\alpha =Z_\Gamma^{-2} Z_Q^{-2} Z_A^{-1} =
1 - \beta_0 \frac{\alpha_{\mathrm{s}}}{4\pi\varepsilon},\quad
\beta_0 = \frac{11}{3} C_A - \frac{4}{3} T_F n_l.
\label{Za}
\end{equation}
This means that the heavy-quark coupling with the gluon field in HQET
is renormalized in the same way as the other QCD couplings.
Of course, this must be the case, because otherwise renormalization
would destroy gauge invariance of HQET.

\subsection{$q$ parallel or orthogonal to $v$}
\label{SubPar}

In the parallel case $q=(\omega'-\omega)v$, $Q=0$.
The denominators of~(\ref{V0def}) are linearly dependent.
Inserting
\[
1 = \frac{(l-q)^2-l^2+2(\omega'-\omega)(l\cdot v+\omega)}
{\omega^{\prime2}-\omega^2}
\]
into the integrand, we obtain
\begin{equation}
\mathcal{V}(\omega,\omega',0) = \frac{1}{\omega'\!+\!\omega}
\left[ 2\mathcal{G}\left((\omega'\!-\!\omega)^2\right)
- \frac{\mathcal{I}(\omega')\!-\!\mathcal{I}(\omega)}{\omega'-\omega} \right],
\end{equation}
exactly at any $d$.

The vertex $\Gamma^\mu$ is, of course, proportional to $v^\mu$.
Therefore, we obtain, either from~(\ref{Gq}) or from~(\ref{Gv}),
\begin{eqnarray}
&& \hspace*{-10mm}
\Gamma_v = 1 - \left(1-\frac{C_A}{2C_F}\right)
\frac{\Sigma(\omega')-\Sigma(\omega)}{\omega'-\omega}
\nonumber\\
&&{} - C_A \frac{g_0^2}{(4\pi)^{d/2}} \frac{\xi}{4(\omega'+\omega)}
\biggl[(d-3)\frac{\omega'\mathcal{I}(\omega')-\omega \mathcal{I}(\omega)}{\omega'-\omega}
\nonumber\\
&&{} + 2\omega\omega'\frac{\mathcal{I}(\omega')-\mathcal{I}(\omega)}{\omega^{\prime2}-\omega^2}
+ \frac{(\omega'-\omega)^2}{\omega'+\omega} \mathcal{G}((\omega'-\omega)^2)
\biggr].
\label{Par}
\end{eqnarray}
It has an imaginary part, which is contained in $\mathcal{G}((\omega'-\omega)^2)$.

The case when $q$ is orthogonal to $v$ ($\omega'=\omega$, $q^2=-Q^2$)
does not lead to great simplifications.
The contribution~(\ref{Vert1a}) of Fig.~\ref{FigVert1}a
now contains $\mathrm{d}\Sigma(\omega)/\mathrm{d}\omega=(d-3)\Sigma(\omega)/\omega$.
The contraction~(\ref{Gq}) is zero,
and hence $\Gamma_b^\mu$ is parallel to $v^\mu$, too.
The contribution of Fig.~\ref{FigVert1}b to $\Gamma_v$
is obtained from~(\ref{Gv}) by putting $\omega'=\omega$.

\subsection{$q$ on the light cone}
\label{SubLC}

When $q^2=0$, the reduction algorithm of Sect.~\ref{SubRec} breaks down.
All $V(\nu_0,\nu_1,\nu_2)$ with $\nu_0\le0$ vanish.
We can use (\ref{r0}) to lower $\nu_0$ down to 1.
Let us suppose that $\nu_2>0$; otherwise, we can interchange
$\nu_1\leftrightarrow\nu_2$, $\omega\leftrightarrow\omega'$.
The relation
\begin{equation}
\bigl[(d-\nu_0-\nu_1-\nu_2)(\omega'-\omega)-\nu_1\omega'+\nu_2\omega
+ \omega \nu_1\1+\2- - \omega'\nu_2\2+\1- \bigr] V = 0,
\hspace*{6mm}
\label{r4}
\end{equation}
which is $\omega'$ times~(\ref{r1}) minus $\omega$ times~(\ref{r2}),
allows us to lower $\nu_2$ down to 1.
We are left with $V(1,\nu_1,1)$.
Taking~(\ref{r2}), subtracting $\omega'$ times~(\ref{r0})
and adding $2\omega'$ times $\1+$ shifted~(\ref{r4}),
we obtain at $\nu_0=\nu_2=1$
\begin{eqnarray}
&&\left[d-\nu_1-3 + 2\omega'(\omega'-\omega)(d-2\nu_1-4)\1+\right] 
V(1,\nu_1,1)
\nonumber\\
&&{}=\left[\nu_1 - 2\omega\omega'(\nu_1+1)\1+\right]\1+\2- V(1,\nu_1,1),
\label{r5}
\end{eqnarray}
where the integrals on the right-hand side are trivial.
Using this relation, we can lower or raise $\nu_1$ to 0.
Therefore, all the integrals $V(\nu_0,\nu_1,\nu_2)$ at $q^2=0$
can be reduced to $\mathcal{I}(\omega)$ and $\mathcal{I}(\omega')$.
For example, for $V(1,1,1)$ we have
\begin{equation}
\mathcal{V}(\omega,\omega',|\omega'-\omega|) =
\frac{(d-3)\bigl(\omega'\mathcal{I}(\omega)-\omega \mathcal{I}(\omega')\bigr)}
{2(d-4)\omega\omega'(\omega'-\omega)},
\label{V0q0}
\end{equation}
at arbitrary $d$.

Repeating, with the new algorithm, the calculation of the diagram
in Fig.~\ref{FigVert1}b at $q^2=0$, we obtain
\begin{eqnarray}
&& \hspace*{-10mm}
\Gamma_b^\mu q_\mu = C_A \frac{g_0^2}{(4\pi)^{d/2}}
\frac{1}{8(d-4)\omega\omega'}
\nonumber\\
&&{}\times\Bigl[2\bigl(d-5+(d-3)(d-4)\xi\bigr)\omega\omega'
\bigl(\mathcal{I}(\omega)-\mathcal{I}(\omega')\bigr)
\nonumber\\
&&\quad{} - (d-3)\bigl(2+(d-4)\xi\bigr)
\bigl(\omega^{\prime2}\mathcal{I}(\omega)-\omega^2 \mathcal{I}(\omega')\bigr)
\Bigr],\hspace{4mm}
\label{Gqq0}\\
&& \hspace*{-10mm}
\Gamma_b^\mu v_\mu = C_A \frac{g_0^2}{(4\pi)^{d/2}}
\frac{(d-3)\xi}{16(d-6)\omega^2\omega^{\prime2}(\omega'-\omega)}
\nonumber\\
&&{}\times\Bigl[ 2(d-6)\omega^2\omega^{\prime2}
\bigl(\mathcal{I}(\omega)-\mathcal{I}(\omega')\bigr)
\nonumber\\
&&\quad{} + \bigl(2-(d-7)\xi\bigr)\omega\omega'
\bigl(\omega^{\prime2}\mathcal{I}(\omega)-\omega^2 \mathcal{I}(\omega')\bigr)
\nonumber\\
&&\quad{} + \bigl(2+(d-5)\xi\bigr)
\bigl(\omega^{\prime4}\mathcal{I}(\omega)-\omega^4 \mathcal{I}(\omega')\bigr) \Bigr].
\label{Gvq0}
\end{eqnarray}
These results can be also obtained from~(\ref{Gq}), (\ref{Gv}),
if we expand the numerator of~(\ref{Gv}) up to the $q^2$ term.

The case $q=0$ belongs to all the categories considered above.
We obtain, from each of the above results, $\Gamma_q=0$,
\begin{eqnarray}
&&\Gamma_v = 1 - \frac{g_0^2}{(4\pi)^{d/2}} (d-3) \frac{\mathcal{I}(\omega)}{8\omega}
\left[ 4 \left( 2+(d-3)\xi \right) C_F
- \left( 4+(d-5)\xi\right) C_A\right],
\hspace*{8mm}
\label{q0}\\
&&\mathcal{V}(\omega,\omega,0) = - \frac{(d-3)\mathcal{I}(\omega)}{2\omega^2},
\nonumber
\end{eqnarray}
exactly at any $d$.

\subsection{$\Omega=0$}
\label{SubOmega}

Another interesting case is $\Omega=0$ ($q^2=-4\omega\omega'$).
After reducing $V(\nu_0,\nu_1,\nu_2)$ to $V(\nu_0,1,1)$ and trivial integrals
(Sect.~\ref{SubRec}), we can use~(\ref{r3}) to reduce $\nu_0$ to 0.
In particular,
\begin{equation}
\mathcal{V}(\omega,\omega',\omega'+\omega) = \frac{d-3}{2(d-4)(\omega'+\omega)}
\left[ 4\mathcal{G}(-4\omega\omega') - \frac{\mathcal{I}(\omega)}{\omega}
- \frac{\mathcal{I}(\omega')}{\omega'} \right],
\hspace*{6mm}
\label{V0Om}
\end{equation}
for any $d$.
Repeating the calculation of the vertex, we obtain
\begin{eqnarray}
&& \hspace*{-10mm}
\Gamma_b^\mu q_\mu = C_A \frac{g_0^2}{(4\pi)^{d/2}}
\frac{1}{16(d-4)(d-6)\omega\omega'(\omega'+\omega)^2} 
\nonumber\\
&&
\times \biggl\{
4 \Bigl[-(d-6)\bigl(4+(d-4)\xi\bigr)(\omega'+\omega)^2
\nonumber\\
&&\quad{}+4(d-3)(d-4)\xi\omega\omega'\Bigr]
\omega\omega'(\omega'-\omega) \mathcal{G}(-4\omega\omega')
\nonumber\\
&&{} + \Bigl[\bigl(4(d-6)+(d-3)(3d-16)\xi\bigr)(\omega'+\omega)^2
\nonumber\\
&&\quad{}-4(d-3)(d-4)\xi\omega\omega'\Bigr]
\nonumber\\
&&\qquad{}\times (d-4) (\omega'+\omega)
\bigl(\omega'\mathcal{I}(\omega)-\omega \mathcal{I}(\omega')\bigr)
\nonumber\\
&&{} + \Bigl[\bigl(4(d-6)-(d-3)(d-4)(d-8)\xi\bigr)(\omega'+\omega)^2
\nonumber\\
&&\quad{}-4(d-3)(d-4)\xi\omega\omega'\Bigr]
\nonumber\\
&&\qquad{}\times (\omega'-\omega)
\bigl(\omega'\mathcal{I}(\omega)+\omega \mathcal{I}(\omega')\bigr) \biggr\},
\label{GqOm}\\
&& \hspace*{-10mm}
\Gamma_b^\mu v_\mu = C_A \frac{g_0^2}{(4\pi)^{d/2}}
\frac{\xi}{32(d-6)(d-8)\omega^2\omega^{\prime2}(\omega'+\omega)^2} 
\nonumber\\
&&
\times \biggl\{
\Bigl[(d-6)(d-8)\bigl(4+(d-4)\xi\bigr)(\omega'+\omega)^4
\nonumber\\
&&\quad{}+4(d-8)\bigl(2d-(d-3)(d-4)\xi\bigr)\omega\omega'(\omega'+\omega)^2
\nonumber\\
&&\quad{}+32(d-3)(d-8-3\xi)\omega^2\omega^{\prime2}\Bigr]
\omega\omega' \mathcal{G}(-4\omega\omega')
\nonumber\\
&&{} - \Bigl[\bigl(2(d-7)(d-8)-(d-4)(d-5)\xi\bigr)(\omega'+\omega)^2
\nonumber\\
&&\quad{}-4(d-5)(d-8-3\xi)\omega\omega'\Bigr]
\nonumber\\
&&\qquad{}\times (d-3) (\omega'+\omega)
\bigl(\omega^{\prime2}\mathcal{I}(\omega)+\omega^2 \mathcal{I}(\omega')\bigr)
\nonumber\\
&&{} + \Bigl[(d-5)\bigl(2(d-8)+(d-4)\xi\bigr)(\omega'+\omega)^2
\nonumber\\
&&\quad{}+4(d-8-3\xi)\omega\omega'\Bigr]
\nonumber\\
&&\qquad{}\times (d-3) (\omega'-\omega)
\bigl(\omega^{\prime2}\mathcal{I}(\omega)-\omega^2 \mathcal{I}(\omega')\bigr) \biggr\}.
\label{GvOm}
\end{eqnarray}

\subsection{Quark(s) on the mass shell}
\label{SubOS}

If one of the quarks is on its mass shell, say $\omega'=0$, then
\begin{equation}
Q \mathcal{V}(\omega,0,Q) = 2 \Li2\left(\frac{Q+\omega}{2\omega}\right)
+ \log^2\frac{Q-\omega}{-2\omega}
 - \frac{1}{2} \log^2\frac{Q-\omega}{Q+\omega}
- \frac{2\pi^2}{3}
\label{V0w0}
\end{equation}
at $d=4$.
The contractions of the vertex are obtained from~(\ref{Gq}), (\ref{Gv})
by setting $\mathcal{I}(\omega')=0$ and then $\omega'=0$.
%\begin{eqnarray}
%&&\Gamma_b^\mu q_\mu = \frac{C_A}{4} \frac{g_0^2}{(4\pi)^{d/2}}
%\Bigl[ 2 \omega^2 \mathcal{V}(\omega,0,Q) - (4-\xi) \omega \mathcal{G}(q^2)
%\nonumber\\
%&&\quad{} + (2+(d-3)\xi) \mathcal{I}(\omega) \Bigr],
%\label{Gqw0}\\
%&&\Gamma_b^\mu v_\mu = C_F \frac{g_0^2}{(4\pi)^{d/2}} \frac{\xi}{8q^2}
%\Bigl[ - 4 (d-4) Q^2 \omega \mathcal{V}(\omega,0,Q)
%\nonumber\\
%&&\quad{} + \bigl((-2(4d-13)+(d-4)\xi)q^2
%\nonumber\\
%&&\qquad{} + (4(2d-7)-(d-4)\xi)\omega^2\bigr) \mathcal{G}(q^2)
%\nonumber\\
%&&\quad{} - 2 (d-3) \omega \mathcal{I}(\omega) \Bigr].
%\label{Gvw0}
%\end{eqnarray}

When both quarks are on shell ($\omega=\omega'=0$),
\begin{eqnarray}
\Gamma_b^\mu q_\mu&=&0,
\nonumber \\
\Gamma_b^\mu v_\mu &=& C_A \frac{g_0^2}{(4\pi)^{d/2}} \frac{\xi}{8}
\left[ - 2 (4d-13) + (d-4)\xi\right] \mathcal{G}(q^2).\hspace{5mm}
\label{Gvw00}
\end{eqnarray}
In this case, $\mathcal{V}(0,0,Q)$ does not appear.

It is easy to consider the cases when $q$ is parallel to $v$ and $\omega'=0$,
and when $q^2=0$, $\omega'=0$.

\section{Conclusion}
\label{Concl}

We have obtained general expressions~(\ref{Vert1a}), (\ref{Gq}) and (\ref{Gv})
for the one-loop HQET quark-gluon vertex. Using recurrence 
relations~(\ref{r0})--(\ref{r3}), we expressed the results in
terms of one non-trivial integral $\mathcal{V}(\omega,\omega',Q)$ (\ref{V0def})
and some trivial integrals (\ref{triv}). For the integral~(\ref{V0def})
in four dimensions, we have obtained an analytic result~(\ref{V0res})
in terms of dilogarithms. In Sections~\ref{SubPar}--\ref{SubOmega}
we have also studied some special limits of interest.

In Appendix~\ref{AppInt} we have provided some results for 
the integrals~(\ref{Vdef})
with arbitrary indices and in arbitrary dimension.
We have also discussed, in Appendix~\ref{AppThreshold}, 
how the HQET limit 
can be obtained directly from the standard integrals occurring 
in the QCD calculation.
Using this prescription, in Appendix~\ref{AppQCD} we have examined 
the $m\to\infty$ limit of the general QCD result~\cite{DOS} 
for the one-loop quark-gluon function, and we have
found that it is in agreement with our calculation. 
We have also presented the result for the background field vertex
(Appendix~\ref{AppBG}).

\vspace{3mm}

%\begin{acknowledgement}
\emph{Acknowledgements.}
We are grateful to V.~A.~Smirnov for useful comments on the
manuscript, and to T.~Mannel and P.~Osland for useful discussions.
A.~G.'s work was supported in part by the RFBR grant 00-02-17646.
A.~D.'s research was supported by the Deutsche Forschungsgemeinschaft.
Partial support 
from RFBR grant 01-02-16171 is also acknowledged.
%\end{acknowledgement}

\appendix

\section{One-loop HQET integrals}
\label{AppInt}

Result for the two-point HQET integral is well known~\cite{BG},
\begin{eqnarray}
\label{def_I}
I(\nu_0, \nu) &\equiv& -\frac{\mathrm{i}}{\pi^{d/2}}
\int \frac{\mathrm{d}^d l}{(l\cdot v+\omega)^{\nu_0}(l^2)^{\nu}}
\nonumber \\
&=& (-1)^{\nu_0+\nu} 2^{\nu_0} (-2\omega)^{d-\nu_0-2\nu}
\frac{\Gamma(\nu_0\!+\!2\nu\!-\!d)\Gamma(d/2\!-\!\nu)}
     {\Gamma(\nu_0)\Gamma(\nu)} .
\end{eqnarray}

For the triangle integral~(\ref{Vdef}), using Feynman parameters, 
we arrive at the following double integral representation:
\begin{eqnarray}
\label{intV}
V(\nu_0,\nu_1,\nu_2) &=&
\frac{(-1)^{\nu_0+\nu_1+\nu_2} 2^{\nu_0}
      \Gamma(\nu_0+\nu_1+\nu_2-d/2)} 
{\Gamma(\nu_0)\; \Gamma(\nu_1)\; \Gamma(\nu_2)}
\nonumber \\ && 
\times \int\limits_0^{\infty} \int\limits_0^{\infty}
\frac{\mathrm{d}y\; \mathrm{d}y'\; y^{\nu_1-1}\; {y'}^{\nu_2-1}
\; (y+y')^{\nu_0+\nu_1+\nu_2-d}}
{\left[1-2\omega y -2\omega' y'
-q^2 y y' \right]^{\nu_0+\nu_1+\nu_2-d/2}} \; .
\end{eqnarray}
The symmetry $(\omega,\nu_1)\leftrightarrow (\omega',\nu_2)$
is explicit.

In the special case $\omega=\omega'=0$, the integral (\ref{intV}) 
can be evaluated in terms of $\Gamma$ functions,
\begin{eqnarray}
\label{special1}
V(\nu_0,\nu_1,\nu_2)\bigl|_{\omega=\omega'=0} &=&
(-1)^{\nu_0+\nu_1+\nu_2} 2^{\nu_0-1}
(-q^2)^{d/2-\nu_0/2-\nu_1-\nu_2}\;
\Gamma(\nu_1+\nu_2+\nu_0/2-d/2)
\nonumber \\ &&
\times
\frac{\Gamma(\nu_0/2) \Gamma(d/2-\nu_0/2-\nu_1)
      \Gamma(d/2-\nu_0/2-\nu_2)}
     {\Gamma(\nu_0) \Gamma(\nu_1) \Gamma(\nu_2)
      \Gamma(d-\nu_0-\nu_1-\nu_2)} \; .
\hspace*{6mm}
\end{eqnarray}
In particular, for $\nu_i=1$ we get
\begin{equation}
\mathcal{V}(0,0,Q)
= - \frac{\Gamma(1/2)\Gamma((5-d)/2)\Gamma^2((d-3)/2)}
{\Gamma(d-3)Q^{1+2\varepsilon}}
= - \frac{\pi^2}{Q} + \mathcal{O}(\varepsilon).
\label{V000}
\end{equation}

In another special case, $q^2=0$, we get a hypergeometric function,
\begin{eqnarray}
\label{q^2=0}
V(\nu_0,\nu_1,\nu_2)\bigl|_{q^2=0} &=&
(-1)^{\nu_0+\nu_1+\nu_2} 2^{\nu_0}
            (-2\omega)^{d-\nu_0-2\nu_1-2\nu_2}
\nonumber \\ && 
\times
\frac{\Gamma(\nu_0+2\nu_1+2\nu_2-d)\;
            \Gamma(n/2-\nu_1-\nu_2)}
{\Gamma(\nu_0)\; \Gamma(\nu_1+\nu_2)}
\nonumber \\ && 
\times
\; _2F_1\biggl( \begin{array}{c}
\nu_2, \; \nu_0+2\nu_1+2\nu_2-d \\ \nu_1+\nu_2 \end{array}
\biggl|~1-\frac{\omega'}{\omega}\biggl) \; .
\end{eqnarray}
Using simple transformations of $_2F_1$ function, it is
easy to see that the result obeys the symmetry
$(\omega,\nu_1)\leftrightarrow (\omega',\nu_2)$, as it should.
Note that the structure of the result (\ref{q^2=0}) is quite
similar to that of the two-point integral with different masses
and zero external momentum, see eq.~(2.9) of \cite{jmp1}.
When $\nu_1=\nu_2=1$, the result (\ref{q^2=0}) reduces to
\begin{eqnarray}
V(\nu_0,1,1)\bigl|_{q^2=0} &=&
(-1)^{\nu_0} 2^{d-4} \Gamma(\nu_0-d+3)\; \Gamma(d/2-2)\;
\nonumber \\ &&
\times
\frac{(-\omega')^{d-\nu_0-3}-(-\omega)^{d-\nu_0-3}}
     {\omega' - \omega} \; .
\end{eqnarray}

Let us represent the denominator of (\ref{intV})
in terms of double Mellin--Barnes integral,
expanding with respect to $\omega$ and $\omega'$ (see, e.g.,
Eq.~(3.4) of~\cite{jmp1}).
Then, the resulting momentum integral can be recognized as
\begin{equation}
\label{shifted}
V(\nu_0+t_1+t_2, \nu_1+t_1, \nu_2+t_2)
\bigl|_{\omega=\omega'=0, \;\; d\to d+2t_1+2t_2} ,
\end{equation}
where $t_1$ and $t_2$ are the contour integration
variables. Using (\ref{special1})
we can evaluate the integral (\ref{shifted})
in terms of $\Gamma$ functions. 
Making a linear substitution for the contour integration
variables ($t_1=s+t$, $t_2=s-t$), we arrive at the following double
Mellin--Barnes representation for the integral (\ref{Vdef}):
\begin{eqnarray}
\label{MBintV} 
V(\nu_0,\nu_1,\nu_2) &=&
\frac{(-1)^{\nu_0+\nu_1+\nu_2} 2^{\nu_0-1}
(-q^2)^{d/2-\nu_0/2-\nu_1-\nu_2}}
{\Gamma(\nu_0)\; \Gamma(\nu_1)\; \Gamma(\nu_2)\;
\Gamma(d-\nu_0-\nu_1-\nu_2)}
\nonumber \\ && 
\times \frac{1}{(2\pi\mathrm{i})^2}
\int\limits_{-\mathrm{i}\infty}^{\mathrm{i}\infty}
\int\limits_{-\mathrm{i}\infty}^{\mathrm{i}\infty}
\mathrm{d}s\; \mathrm{d}t\; \left(-\frac{4\omega\omega'}{q^2}\right)^s
\left(\frac{\omega}{\omega'}\right)^t \;
\Gamma(-s-t)\; \Gamma(t-s)
\nonumber \\ && 
\times
\Gamma(d/2-\nu_0/2-\nu_1-t)\;
\Gamma(d/2-\nu_0/2-\nu_2+t)
\nonumber \\ && 
\times
\Gamma(\nu_0/2+\nu_1+\nu_2-d/2+s)\;
\Gamma(\nu_0/2+s) \; .
\end{eqnarray}

\section{QCD integrals in the HQET limit}
\label{AppThreshold}

Here we discuss the relation of massive integrals
occurring in standard QCD calculations
and HQET integrals.
First, let us consider the two-point integral with 
one massive line and one massless line,
\begin{equation}
\label{def_J}
J(\nu_0,\nu;m,0) = \int
\frac{\mathrm{d}^d l}{\left[(p+l)^2-m^2\right]^{\nu_0} (l^2)^\nu}.
\label{Jdef}
\end{equation}
For general values of $\nu_0$, $\nu$ and $d$, such integrals
have been examined in~\cite{BD-TMF,GJT}.

When we substitute $p=mv+k$, the massive denominator in (\ref{def_J}) becomes 
\begin{equation}
\label{heavy_prop}
\left[2m(l+k)\cdot v+(l+k)^2\right]^{-\nu_0} \; .
\end{equation}
For $k\ll m$, several regions of integration in $l$ are essential.
When $l\sim k$, we can expand the heavy propagator (\ref{heavy_prop})
in both $k/m$ and $l/m$, and it becomes the HQET propagator.
The leading term of this HQET contribution (called ``ultrasoft'' 
in~\cite{BS,Smi-PLB})
yields $1/m^{\nu_0}$ times an HQET integral (\ref{def_I}), which is proportional to
$(-2\omega)^{d-\nu_0-2\nu}$, by dimensionality.
Higher terms form an expansion in $k/m$.

Let us subtract and add this expansion of the heavy propagator to the exact one.
In the difference, the contribution of small $l\sim k$ is suppressed;
typically, $l\sim m$.
Therefore, we can expand this integrand difference in regular series
in $k/m$, and integrate term by term.
Integrals of all terms of the HQET integrand expansion in $k/m$
vanish in dimensional regularization, because they contain no scale.
Therefore, this ``hard'' contribution can be obtained by expanding the exact
QCD integrand~(\ref{heavy_prop}) in $k/m$, and integrating term by term.
It is analytical at $k=0$, by construction.
The leading term is proportional to $m^{d-2\nu_0-2\nu}$,
by dimensionality, whereas the
higher terms form an expansion in $k/m$.
This separation of $J(\nu_0,\nu;m,0)$ at $k\ll m$ into two 
contributions~\cite{GJT}
is a particular case of a more general threshold expansion~\cite{BS}.
Note that $k$ plays a role of the threshold parameter, 
since $p^2-m^2\sim mk$.

We can check these qualitative considerations, using an explicit expression
for $J(\nu_0,\nu;m,0)$.
It can be presented in terms of $_2F_1$ function of $z=p^2/m^2$
(see Eq.~(10) of~\cite{BD-TMF}).
Note that in the HQET limit $z$ approaches 1,
\[
z=\frac{(mv+k)^2}{m^2}, \quad 1-z=-\frac{2\omega}{m}-\frac{k^2}{m^2} \; ,
\]
i.e., it is at the border of convergence of the $_2F_1$ function.
Transforming from the variable $z$ to $1-z$, we obtain
(see also Eqs.~(1.12)--(1.15) of~\cite{GJT})
\begin{eqnarray}
J(\nu_0,\nu;m,0) &=& \mathrm{i} \pi^{d/2} (-1)^{\nu_0+\nu}
m^{d-2\nu_0-2\nu}
\nonumber \\ &&
\times \Biggl\{ 
\frac{\Gamma(\nu_0+\nu-d/2)\Gamma(d-\nu_0-2\nu)}
     {\Gamma(\nu_0)\Gamma(d-\nu_0-\nu)}\;
_2F_1\biggl( \begin{array}{c} \nu,\; \nu_0+\nu-d/2 \\
             \nu_0+2\nu-d+1 \end{array}
\biggl|~1-z\biggl)
\nonumber \\ &&
+ \frac{\Gamma(\nu_0+2\nu-d)\Gamma(d/2-\nu)}
     {\Gamma(\nu_0)\Gamma(\nu)}\;
(1-z)^{d-\nu_0-2\nu}
\nonumber \\ &&
\times 
_2F_1\biggl( \begin{array}{c} d/2-\nu,\; d-\nu_0-\nu \\
             d-\nu_0-2\nu+1 \end{array}
\biggl|~1-z\biggl)
\Biggl\} \; .
\end{eqnarray}

We see that the first term here is nothing but the ``hard'' contribution.
It has a prefactor $m^{d-2\nu_0-2\nu}$.
The prefactor of the second $_2F_1$ function is
\[
m^{d-2\nu_0-2\nu} (1-z)^{d-\nu_0-2\nu}
\Rightarrow m^{-\nu_0} (-2\omega)^{d-\nu_0-2\nu} ,
\] 
up to higher powers of $1/m$.
This is the HQET (``ultrasoft'') contribution;
in the leading order, it yields 
\[
J(\nu_0, \nu; m, 0) \Rightarrow
{\mathrm i} \pi^{d/2} (2m)^{-\nu_0} I(\nu_0,\nu),
\]
where $I(\nu_0,\nu)$ is defined in~(\ref{def_I}).

The value of $(2m)^{\nu_0}J(\nu_0,\nu;m,0)$ at the singular point $1/m=0$ is
(up to a factor $\mathrm{i}\pi^{d/2}$) 
the HQET integral~(\ref{def_I}).
When setting $1/m=0$, we discard the ``hard'' contribution,
which is proportional to $m^{d-\nu_0-2\nu}$,
because we can always choose $d$ small enough
for this contribution to vanish.
The naive Taylor expansion in $1/m$ is given by the HQET contribution.
Similarly, the value of $J(\nu_0,\nu;m,0)$ at the singular point $k=0$ is the
QCD on-shell integral.
When setting $k=0$, we discard the HQET contribution,
which is proportional to $(-2\omega)^{d-\nu_0-2\nu}$,
because we can always choose $d$ large enough
for this contribution to vanish.
The naive Taylor expansion in $k$ is given by the ``hard'' contribution.
Neither of these naive expansions, taken separately, describes
$J(\nu_0,\nu;m,0)$ at $k\ll l$.
It is given by the sum of both contributions, ``hard'' and HQET
(``ultrasoft'') ones (see in~\cite{BS}).

Now, let us consider off-shell three-point integrals with one
massive and two massless internal lines.
For arbitrary powers of propagators, such vertex
integrals has been considered in~\cite{BD-TMF} (see Eqs.~(25)--(29))
and~\cite{jmp2} (Eqs.~(4.1)--(4.5)).
Using the Mellin--Barnes representation,
Eq.~(4.2) of~\cite{jmp2} (see also Eq.~(26) of~\cite{BD-TMF}),
we see that two (of three) arguments, $p_1^2/m^2$ and $p_2^2/m^2$
(we denote $p_1^2=k_{23}^2$, $p_2^2=k_{13}^2$, $p_3^2=k_{12}^2$)
are approaching 1 in our limit, which is at the border
of convergence of the corresponding function.

Again, as in the two-point case, we need to construct analytic continuation  
of this representation, in terms of the variables
containing $(m^2-p_1^2)$ and $(m^2-p_2^2)$. To do this inside
the Mellin--Barnes representation, we use the
contour-integral version of the corresponding formula
for $_2F_1$, namely, Eq.~(A7) of~\cite{jmp2} (where we put $\nu_3=\nu_0$).
In this way, we arrive at
\begin{eqnarray}
J_1(\nu_1,\nu_2,\nu_0; m) &=& \frac{\mathrm{i} \pi^{d/2}\; (-1)^{\nu_0+\nu_1+\nu_2}
 m^{d-2\nu_0-2\nu_1-2\nu_2}}
{\Gamma(\nu_1) \Gamma(\nu_2) \Gamma(\nu_0)
       \Gamma(d\!-\!\nu_1\!-\!\nu_2\!-\!\nu_0)}
\nonumber \\ &&
\times  \frac{1}{(2\pi \mathrm{i})^3} \!
\int\!\!\!\!\int\limits_{-\mathrm{i}\infty}^{\mathrm{i}\infty}\!\!\!\!\int \!
\mathrm{d}s \; \mathrm{d}t \; \mathrm{d}u
\biggl( \!\frac{m^2\!-\!p_1^2}{m^2} \!\biggl)^{\! s} \!
\biggl( \!\frac{m^2\!-\!p_2^2}{m^2} \!\biggl)^{\! t} \!
\biggl( -\frac{p_3^2}{m^2} \biggl)^{\! u} \!
\nonumber \\ &&
\times \Gamma(-s) \Gamma(-t) \Gamma(-u)
\Gamma(d\!-\!\nu_0\!-\!2\nu_1\!-\!2\nu_2\!-\!s\!-\!t\!-\!2u)
\nonumber \\ &&
\times \Gamma(\nu_0+\nu_1+\nu_2-d/2+s+t+u)\;
\nonumber \\ &&
\times \Gamma(\nu_1+t+u)\; \Gamma(\nu_2+s+u) \; .
\end{eqnarray}

Now, we need to analyse the contributions of the poles in the
right half-plane of the contour variable $u$, since they
correspond to increasing powers of $p_3^2/m^2$.
There are only two series of poles (with $j=0,1,2,\ldots$), 
\begin{eqnarray*}
\mbox{(i)}\;\; u&=&j\;\; (\mbox{due to}\;\; \Gamma(-u))\;\; 
\mbox{and} 
\\
\mbox{(ii)}\;\; u&=&d/2-\nu_0/2-\nu_1-\nu_2-s/2-t/2+j/2
\\ &&
(\mbox{due to}\;\; \Gamma(d-\nu_0-2\nu_1-2\nu_2-s-t-2u)). 
\end{eqnarray*}
The prefactors contain $m^{d-2\nu_0-2\nu_1-2\nu_2}$ (case (i))
and $m^{-\nu_0}$ (case (ii)).
The series~(i) yields the ``hard'' contribution.
The HQET (``ultrasoft'') contribution is given by the series~(ii),
whose leading term is $j=0$.
These two contributions are related to the singular limits
$1/m=0$ and $k=q=0$ in a way similar to the two-point case.

Picking up this leading HQET contribution, we arrive at a double
Mellin--Barnes representation (in terms of the remaining
contour integrals over $s$ and $t$),
where the kinematical variables involved
yield (in the limit $m\to\infty$) the HQET variables,
\[
\frac{m^2-p_1^2}{m\sqrt{-p_3^2}} \to \frac{-2\omega'}{\sqrt{-q^2}},
\quad
\frac{m^2-p_2^2}{m\sqrt{-p_3^2}} \to \frac{-2\omega}{\sqrt{-q^2}} .
\]
After introducing the new contour 
variables $(s+t)/2$ and $(s-t)/2$
we see that
the resulting Mellin--Barnes representation is equivalent
to (\ref{MBintV}), so that
\[
J_1(\nu_1,\nu_2,\nu_0;m) \Rightarrow \mathrm{i} \pi^{d/2} 
(2m)^{-\nu_0} V(\nu_0, \nu_1, \nu_2) \; .
\]

Therefore, the HQET integrals can be formally obtained \emph{directly} from
the standard loop integrals, by picking up the formal Taylor series 
in $1/m$ which has no prefactors containing $m$ to the power
depending on $d$. Such prescription is similar to some other
prescriptions in dimensional regularization. For instance, 
considering the massless limit of the integral~(\ref{def_J})
we need to represent the result in terms of the functions of
the variable $m^2/p^2$, and then discard the contribution containing
$m^{-2\varepsilon}$ (see Eq.~(11) of~\cite{BD-TMF}).  
We have also demonstrated that the HQET contributions are
equivalent to the ``ultrasoft'' ones, in the language of the threshold
expansion~\cite{BS}. 
In other words, in the cases considered the
exact result is given by the sum of two contributions.
The first one is given by a formal Taylor expansion of the integrand
in the small parameter of the threshold expansion (the ``hard''
contribution). The second one is nothing but the HQET series
in $1/m$.  

\section{QCD vertex at $m\to\infty$}
\label{AppQCD}

If we substitute the ``hard'' parts of all scalar integrals into the QCD vertex,
then, in the limit $k\to0$, $q\to0$, we just get the on-shell vertex at $q=0$.
Corrections to this limit are regular expansion terms in $k/m$, $q/m$.
Since we do not consider $1/m$ suppressed terms here,
these corrections can be omitted.

If we substitute the HQET parts of all scalar integrals 
(see Appendix~\ref{AppThreshold}),
we should obtain the HQET vertex, which was calculated in Sect.~\ref{SecVert}.
In order to make a strong check of both the results of~\cite{DOS}
(where the one-loop quark-gluon vertex was calculated in arbitrary
gauge and dimension)
and of the present ones, we consider here the HQET limit 
of the QCD vertex~\cite{DOS}.

Using the standard decomposition of the quark-gluon vertex~\cite{BC1}
(see also in~\cite{KRP,DOS}), it can be split into longitudinal
and transverse parts,
\begin{equation}\
\label{BC}
\Gamma^{\mu} = \sum_{i=1}^4 \lambda_i L_i^{\mu} + 
\sum_{i=1}^8 \tau_i T_i^{\mu} \; ,
\end{equation}
where $\lambda_i$ and $\tau_i$ are scalar functions depending on
kinematical variables, whereas $L_i^{\mu}$ and $T_i^{\mu}$ are
vectors which also involve Dirac matrices (see Sect.~IID of~\cite{DOS} 
for further details). 

We substitute
\[
p_1 = - mv - k - q,\quad
p_2 = mv + k,\quad
p_3 = q.
\]
At the leading order in $1/m$, the initial and final quark spinors
obey $\rlap/v u=u$.
Therefore, we can sandwich $\Gamma^\mu$ between the projectors
$(\rlap/v+1)/2$, and use
\[
\frac{\rlap/v+1}{2} \gamma^\mu \frac{\rlap/v+1}{2} =
\frac{\rlap/v+1}{2} v^\mu \frac{\rlap/v+1}{2}.
\]
With the required accuracy, there are three independent structures:
\begin{eqnarray*}
&&L_1^\mu \Rightarrow v^\mu + \mathcal{O}(1/m),\\
&&T_3^\mu \Rightarrow q^2 v^\mu - (\omega'-\omega) q^\mu + \mathcal{O}(1/m),\quad
T_5^\mu \Rightarrow \frac{1}{2} [\gamma^\mu,\rlap/q].
\end{eqnarray*}
The others are
\begin{eqnarray*}
&&L_2^\mu \Rightarrow 4 m^2 L_1^\mu + \mathcal{O}(m),\quad
L_3^\mu \Rightarrow - 2 m L_1^\mu + \mathcal{O}(1),\quad
L_4^\mu \Rightarrow \mathcal{O}(1),\\
&&T_1^\mu \Rightarrow m T_3^\mu + \mathcal{O}(1),\quad
T_2^\mu \Rightarrow 2 m^2 T_3^\mu + \mathcal{O}(m),\\
&&T_4^\mu \Rightarrow \mathcal{O}(m),\quad
T_6^\mu \Rightarrow \mathcal{O}(1),\quad
T_7^\mu \Rightarrow \mathcal{O}(m),\quad
T_8^\mu \Rightarrow m T_5^\mu + \mathcal{O}(1).
\end{eqnarray*}
According to Appendix~\ref{AppThreshold}, the HQET-relevant
``ultrasoft'' parts of the scalar integrals are
\begin{eqnarray*}
&&\eta\tilde{\kappa} \Rightarrow 0,\quad
\eta\kappa_{2,3} \Rightarrow 0,\quad
\eta\kappa_{0,3} \Rightarrow \mathcal{G}(q^2),\\
&&\eta\kappa_{1,2} \Rightarrow \frac{\mathcal{I}(\omega)}{2m},\quad
\eta\kappa_{1,1} \Rightarrow \frac{\mathcal{I}(\omega')}{2m},\\
&&\eta\varphi_1 \Rightarrow \frac{\mathcal{V}(\omega,\omega',Q)}{2m},\quad
\eta\varphi_2 \Rightarrow \frac{1}{(2m)^2}
\frac{\mathcal{I}(\omega')-\mathcal{I}(\omega)}{\omega'-\omega},
\end{eqnarray*}
where the notations $\varphi_i$, $\kappa_{i,l}$ and $\eta$ are 
defined in Sect.~IIA of~\cite{DOS}.

Therefore, at the leading order ($1/m^0$) the HQET limit of the QCD vertex
(\ref{BC}) yields
\[
(\lambda_1+4m^2\lambda_2-2m\lambda_3) L_1^{\mu}
+(m\tau_1+2m^2\tau_2+\tau_3) T_3^{\mu} 
+(\tau_5+m\tau_8) T_5^{\mu} \; . 
\]
Using the results for $\lambda_i$ and $\tau_i$ listed in~\cite{DOS},
we have obtained that
at the order $1/m^0$ the coefficient of the chromomagnetic structure $T_5^\mu$
vanishes, whereas
those of $L_1^\mu$ and $T_3^\mu$ reproduce the results 
(\ref{Vert1a}), (\ref{Gq}) and (\ref{Gv}),
for arbitrary $d$ and $\xi$.

Thus the QCD vertex at small $k$, $q$
is equal to its on-shell value at $k=q=0$ plus the HQET vertex,
up to $1/m$ corrections.
We can reformulate this statement:
the QCD vertex in the on-shell renormalization scheme
is equal to the HQET vertex in the on-shell renormalization scheme,
up to $1/m$ corrections.
In QCD, the on-shell renormalization subtracts from the one-loop correction
its value at $k=q=0$.
In HQET, the on-shell renormalized vertex equals the bare one,
because its value at $k=q=0$ vanishes.

\section{Background-field vertex}
\label{AppBG}

The background field formalism~\cite{Ab} is convenient for considering
heavy quark scattering in an external gluon field.
In this method, the vertex (Fig.~\ref{FigVert}) with the background gluon
differs from the ordinary one.
The diagram of Fig.~\ref{FigVert1}a still gives~(\ref{Vert1a}),
because the quark-gluon elementary vertex does not change.
The contraction of the three-gluon vertex with a background gluon momentum
contains no ghost terms,
just the first difference in Fig.~\ref{FigWardQG}b.
Therefore,
\[
\Gamma_b^\mu q_\mu = \frac{C_A}{2 C_F}
\left(\Sigma(\omega)-\Sigma(\omega')\right).
\]
This is also confirmed by a direct calculation.
For the other contraction, we obtain
\begin{eqnarray*}
&& \hspace*{-10mm}
\Gamma_b^\mu v_\mu = C_A \frac{g_0^2}{(4\pi)^{d/2}}
\frac{1}{8 q^2 \Omega^2}\\
&&{}\times\biggl\{ 2 (\omega'+\omega) \Bigl[ 2 \Omega^2 q^2
- (d-8)\xi\Omega^2 Q^2 - d\xi\Omega Q^2 q^2\\
&&\qquad{} - (16+(d^2-18d+40)\xi)\xi\Omega Q^2\omega\omega'\\
&&\qquad{} - (d-3)(d-4)\xi^2\Omega q^2\omega\omega'
- d^2\xi^2 Q^2 q^2\omega\omega'\\
&&\qquad{} - 8(5d-12)\xi^2 Q^2 \omega^2\omega^{\prime2}
\Bigr] \mathcal{V}(\omega,\omega',Q)\\
&&\quad{} + \Bigl[ - 8 \Omega^2 q^2
+ (4(2d-7)-(d-4)\xi)\xi\Omega^2 Q^2\\
&&\qquad{} - 8(d-3)(2-(d-5)\xi)\xi\Omega Q^2\omega\omega'\\
&&\qquad{} + 4(d-3)(d-4)\xi^2\Omega q^2\omega\omega'\\
&&\qquad{} - 16(d-3)(d-6)\xi^2 Q^2\omega^2\omega^{\prime2} \Bigr] \mathcal{G}(q^2)\\
&&\quad{} + 2 (d-3) \xi \Bigl[
(2+\xi)q^2\bigl(\omega^{\prime3}\mathcal{I}(\omega)+\omega^3 \mathcal{I}(\omega')\bigr)\\
&&\qquad{} + 2(4+(d-4)\xi)\omega\omega'(\omega^{\prime2}-\omega^2)
\bigl(\omega'\mathcal{I}(\omega)-\omega \mathcal{I}(\omega')\bigr)\\
&&\qquad{} - \bigl((1+\xi)\Omega+(d-6)\xi\omega\omega'\bigr)q^2
\bigl(\omega'\mathcal{I}(\omega)+\omega \mathcal{I}(\omega')\bigr)\\
&&\qquad{} - \bigl(q^2+(6+(d\!-\!5)\xi)\omega\omega'\bigr)q^2
\bigl(\omega \mathcal{I}(\omega)+\omega'\mathcal{I}(\omega')\bigr)\Bigl]\biggl\}.
\end{eqnarray*}

\end{document}